\documentclass[aps,prb,reprint,showpacs,superscriptaddress,floatfix]{revtex4-2}

\usepackage{graphicx}
\usepackage{amsmath,amssymb,amsfonts}
\usepackage[normalem]{ulem}           

\begin{document}


\title{Broken inversion symmetry in
the charge density wave phase in EuAl$_4$}

\author{Surya Rohith Kotla}
\affiliation{Laboratory of Crystallography,
Bayerisches Geoinstitut,
University of Bayreuth, 95440 Bayreuth, Germany}

\author{Leila Noohinejad}
\affiliation{Deutsches Elektronen-Synchrotron DESY,
Notkestrasse 85, 22607 Hamburg, Germany}

\author{Preeti Pokhriyal}
\affiliation{Deutsches Elektronen-Synchrotron DESY,
Notkestrasse 85, 22607 Hamburg, Germany}

\author{Martin Tolkiehn}
\affiliation{Deutsches Elektronen-Synchrotron DESY,
Notkestrasse 85, 22607 Hamburg, Germany}

\author{Harshit Agarwal}
\affiliation{Laboratory of Crystallography,
Bayerisches Geoinstitut,
University of Bayreuth, 95440 Bayreuth, Germany}
\affiliation{Institut f\"{u}r Physik,
Johannes-Gutenberg-Universit\"{a}t Mainz,
55128 Mainz, Germany}

\author{Sitaram Ramakrishnan}
\affiliation{Institut N\'{e}el CNRS/UGA UPR2940,
25 Rue des Martyrs, 38042 Grenoble, France}

\author{Sander van Smaalen}
\email{smash@uni-bayreuth.de}
\affiliation{Laboratory of Crystallography,
Bayerisches Geoinstitut,
University of Bayreuth, 95440 Bayreuth, Germany}

\date{\today}

\begin{abstract}
EuAl$_4$ exhibits a complex phase diagram,
including the development of a charge density wave (CDW)
below $T_{CDW} = 145$ K.
Below $T_{N}=15.4$ K, a series of antiferromagnetically
(AFM) ordered phases appear, while non-trivial
topological phases, like skyrmion lattices, are
stabilized under an applied magnetic field.
The symmetries of the variously ordered phases
are a major issue concerning the understanding of
the stabilization of the ordered phases
as well as concerning the interplay between the
various types of order.
EuAl$_4$ at room temperature
has tetragonal symmetry with space group $I4/mmm$.
The CDW phase has an incommensurately modulated
crystal structure described by
the modulation wave vector
$\mathbf{q} \approx 0.17\,\mathbf{c}^{*}$.
On the basis of various experiments,
including elastic and inelastic x-ray scattering,
and second-harmonic generation,
it has been proposed that the symmetry
of the CDW phase of EuAl$_4$ could be
centrosymmetric orthorhombic, non-centrosymmetric
orthorhombic or non-centrosymmetric tetragonal.
Here, we report temperature-dependent,
single-crystal x-ray diffraction experiments
that show that the CDW is a transverse CDW
with phason disorder, and
with non-centrosymmetric symmetry according to
the orthorhombic superspace group $F222(0\,0\,\sigma)00s$.
Essential for this finding is the availability
of a sufficient number of second-order ($2\mathbf{q}$)
satellite reflections in the x-ray diffraction data set.
The broken inversion symmetry implies that skyrmions
might form due to Dzyaloshinskii-Moriya (DM) interactions,
instead of a more exotic mechanism as it is required for
centrosymmetric structures.
\end{abstract}

\maketitle

\section{\label{sec:eual4_introduction}Introduction}

The charge density wave (CDW) was originally proposed
as an instability of quasi-one-dimensional (1D) metals
that is stabilized by Fermi-surface nesting (FSN)
\cite{gruener1994a,monceau2012a}.
Since then, Materials have been discovered that
possess a CDW at low temperatures, but that have a
three-dimensional (3D) electronic band structure.
Stabilization of these CDWs is by alternate mechanisms,
for example by momentum-dependent electron-phonon
coupling (EPC) instead of FSN \cite{pougetjp2024a}.
A CDW leads to a modified band structure as well as
displacements of the atoms out of their lattice-periodic
positions.
Both the electron density and the atomic displacements
are modulated in a wave-like manner.
These waves are characterized by a common modulation
wave vector $\mathbf{q}$, that is in general
incommensurate with the underlying lattice.
These features imply an interaction or competition
between CDWs and other electronic
properties \cite{khomskii2021a}.
For example, superconductivity (SC) is enhanced
upon suppression of the CDW by application of
pressure or chemical doping, as it has been found for
Lu$_5$Ir$_4$Si$_{10}$ \cite{yanghd1991b,ramakrishnan2017a},
and the kagome lattice compound
CsV$_3$Sb$_5$ \cite{wang2021c,xia2022a}.
Several compounds $R$NiC$_2$ ($R$ = rare earth) possess
a CDW, while at lower temperatures magnetic order develops.
The CDW coexists with antiferromagnetic (AFM) order
\cite{kolincio2017a,roman2023a}, while the CDW
disappears on entering the ferromagnetic (FM) state
\cite{shimomura2009a,wolfel2010a}.
On the other hand, magnetic order is suppressed by
the presence of a CDW in Er$_2$Ir$_3$Si$_5$
\cite{ramakrishnan2020a,singhy2004a}.

The topological magnets EuAl$_4$ and
EuAl$_2$Ga$_2$ \cite{moyajm2022a,vibhakar2023a}
have attracted attention due to the presence of
CDWs and magnetic order at low temperatures
\cite{takagir2022a,gen2023a,yangr2024a}.
The complex phase diagrams include four
differently ordered magnetic phases below
$T_{N}=15.4$ K in EuAl$_4$ as well as
several metamagnetic phases, including two
types of skyrmion lattices under applied
magnetic field \cite{takagir2022a,gen2023a}.
Magnetic order develops out of the CDW phase.
The symmetry of the CDW thus is of high
importance for understanding
the microscopic mechanism of magnetic order.

\begin{figure}
\includegraphics[width=78mm]{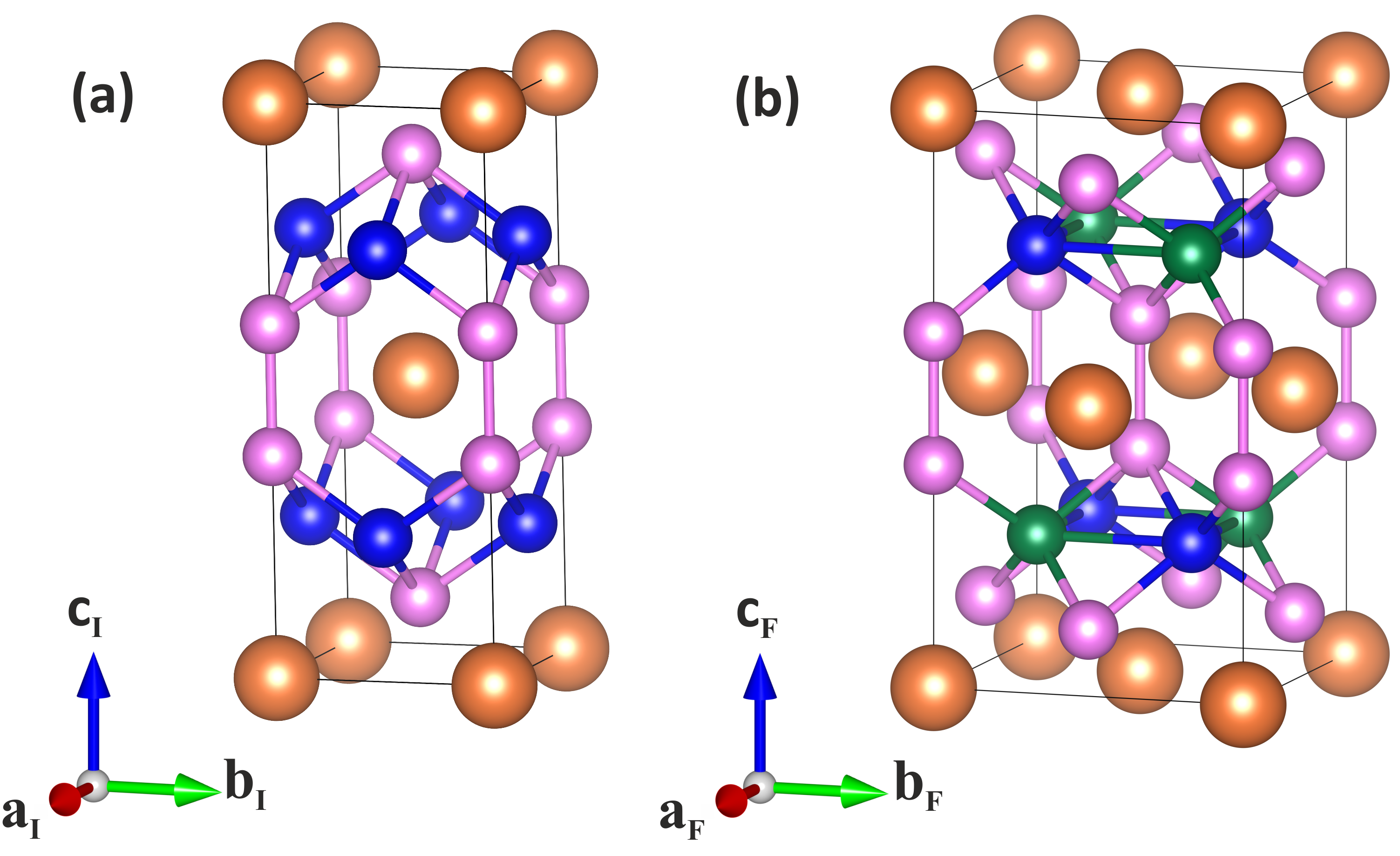}
\caption{\label{fig:eual4_basic_unit_cell}%
(a) Crystal structure of EuAl$_4$ with
space group $I4/mmm$ in the periodic phase at 160 K.
Depicted is an $I$-centered unit cell with basis vectors
$\mathbf{a}_I$, $\mathbf{b}_I$ and $\mathbf{c}_I$.
(b) Basic structure of EuAl$_4$ in the CDW phase at 30 K,
showing an $F$ centered unit cell with the with basis vectors
$\mathbf{a}_F$, $\mathbf{b}_F$ and $\mathbf{c}_F$
and space group $F222$.
The relation between the $I$-centered and $F$-centered
unit cells is:
$\mathbf{a}_F = (\mathbf{a}_I + \mathbf{b}_I)$,
$\mathbf{b}_F = (-\mathbf{a}_I + \mathbf{b}_I)$
and $\mathbf{c}_F = \mathbf{c}_I$.
Orange spheres correspond to the Eu atoms;
blue spheres represent Al1/Al1a atoms;
green spheres represent Al1b atoms;
and pink spheres stand for Al2 atoms.
}
\end{figure}

At room temperature, EuAl$_4$ has the BaAl$_4$
structure type with tetragonal symmetry $I4/mmm$
(Fig. \ref{fig:eual4_basic_unit_cell}).
Originally, it was proposed that the symmetry
of EuAl$_4$ would remain tetragonal in its CDW
phase, because no lattice distortions could be
observed \cite{shimomura2019a}.
While this observation has been confirmed in
several studies by x-ray diffraction,
including the present study,
lower symmetries were proposed for the CDW phase,
including the orthorhombic superspace groups
$Fmmm(0\,0\,\sigma)s00$ and $Immm(0\,0\,\sigma)s00$
\cite{ramakrishnan2022a,korshunov2024a,sukhanovas2025a,agarwalh2025a}.
The tetragonal point groups $4$ and $4/m$ were
suggested on the basis of second-harmonic-generation
(SHG) measurements \cite{yangfz2024a}.

Here, we report the incommensurately modulated
crystal structure of the CDW state of EuAl$_4$
with the non-centrosymmetric orthorhombic symmetry
$F222(0\,0\,\sigma)00s$, as obtained from
temperature-dependent single-crystal x-ray diffraction (SXRD).
The non-centrosymmetric symmetry can only be
distinguished from the previously suggested
centrosymmetric superspace groups through the
inclusion of second-order $(m=2)$ satellite
reflections in the structural analysis.
Second-order satellites were not available in previous studies
\cite{ramakrishnan2022a,korshunov2024a}.
Furthermore, the structural analysis indicates
phason disorder in the CDW \cite{perezmato1991a}.

The present result corroborates the analysis from
cryogenic four-dimensional scanning transmission
electron microscopy (4D-STEM), on the basis of which
a structural modulation was proposed without
assigning a symmetry to it \cite{nih2024a}.
It also mirrors the symmetry obtained for the
isostructural, non-magnetic CDW compound SrAl$_4$,
for which the SXRD data include second-order
satellites \cite{ramakrishnan2024a}.

The proposed symmetry is important for the description
of the magnetically ordered phases.
The non-centrosymmetric superspace group allows
for a stabilization of skyrmions by
Dzyaloshinskii-Moriya (DM) interactions,
the latter which are usually the source of skyrmions
\cite{muhlbaue2009a,nagaosa2013a,tokur2021a}.
A more exotic mechanism, like those involving
itinerant-electron-mediated interactions,
is not required \cite{takagir2022a,gen2023a,nih2024a}.
Secondly, Vibakhar \textit{et.al} \cite{vibhakar2024b}
have found that the magnetic symmetry becomes polar monoclinic
at the onset of the third AFM phase, while the symmetry
of the CDW would lower to point group $222$ or $2$.
The present result of the non-centrosymmetric superspace
symmetry $F222(0\,0\,\sigma)00s$ for the CDW phase
demonstrate that point group $222$ has already been reached,
before any magnetic order develops.

\section{\label{sec:eual4_experiment}Experiment}

\subsection{\label{sec:eual4_crystal_growth}Crystal growth}

\begin{figure}
\includegraphics[width=70mm]{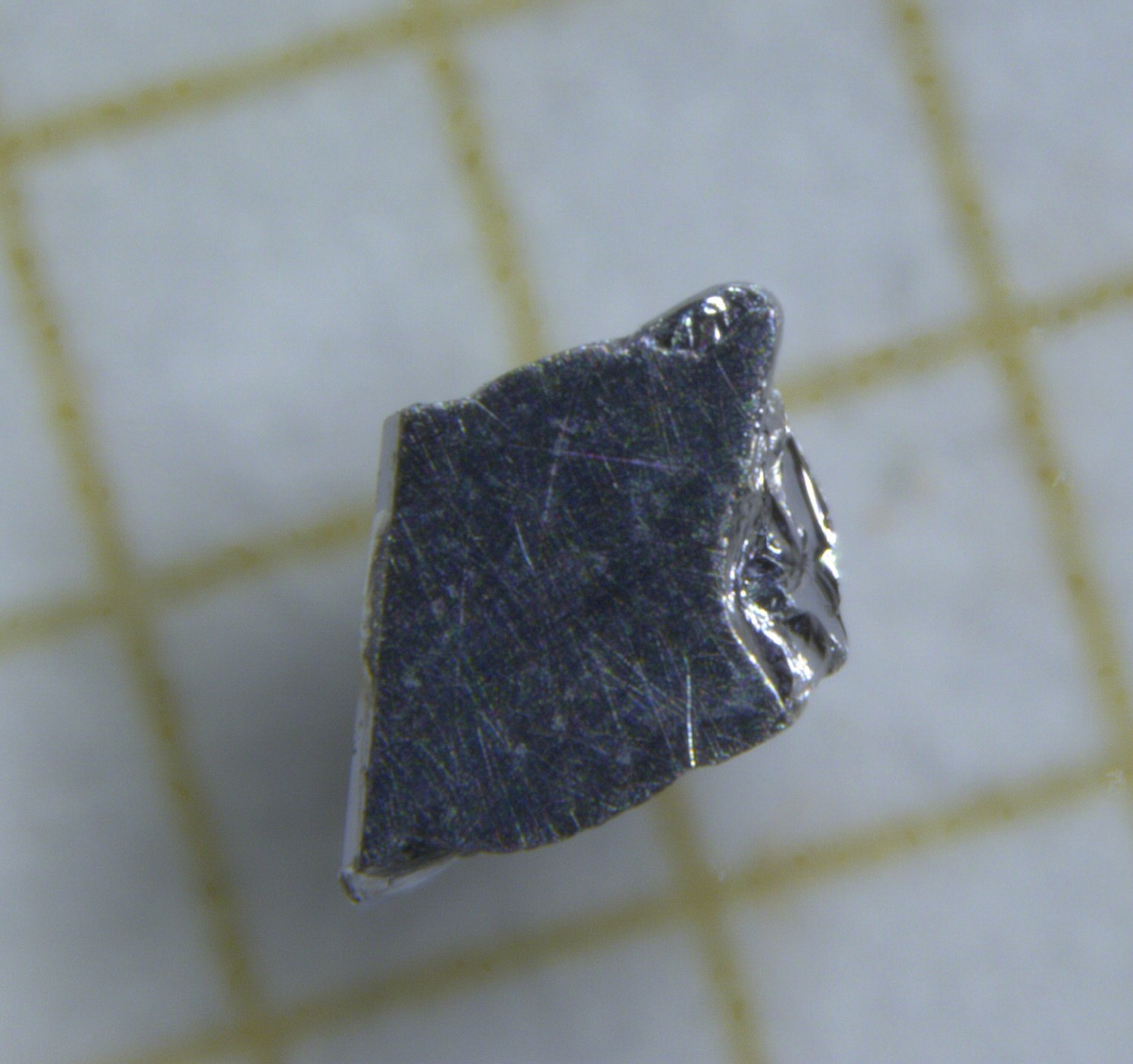}
\caption{\label{fig:eual4_crystal_foto}%
Single crystal of EuAl$_4$ as separated from the Al flux.
Yellow lines form a mesh of $1\times 1$ mm$^{2}$.
}
\end{figure}

High-quality single crystals of EuAl$_4$ were grown
by the self-flux method according to the procedure
described by Nakamura \textit{et al.} \cite{nakamura2015a},
using growth parameters different from our previous
work \cite{ramakrishnan2022a}.
Europium (Smart-elements, 99.99\% purity) and
aluminum (Alfa-Aesar, 99.9995\%)
were mixed into an alumina crucible in the elemental
ratio $1\: :\:9$, and then sealed under vacuum
in a quartz-glass tube.
The quartz-glass tube was heated to 1173 K and
maintained at this temperature for two days.
It was then slowly cooled down to 673 K over 110 hours.
The oven was tilted by 9 deg, with the materials
at the lower side, thus keeping all materials together
during the reaction.
After cooling to room temperature, a single crystal
was obtained of about $1$ mm in size
(Fig. \ref{fig:eual4_crystal_foto}).

We believe that the present single crystal was of
better quality (fewer lattice defects) than the crystal
used in our earlier study \cite{ramakrishnan2022a}.
Most likely, this is due to an effective annealing
of the present material, as it has resulted from
the slow cooling down to 673 K,
as opposed to quenching from 923 K to room temperature
in \cite{ramakrishnan2022a}.

\subsection{\label{sec:eual4_xray_diffraction}X-ray diffraction}

The as-grown single crystal was crushed,
and a small part of dimensions
$0.063 \times 0.041\times 0.005$ mm$^3$
was selected for single-crystal X-ray diffraction (SXRD).
Diffraction experiments were performed with synchrotron
radiation at Beamline P24 of PETRA III at DESY in
Hamburg, Germany, employing radiation of a wavelength
of $0.5000$~\AA{}.
A CRYOCOOL G2B-LT open-flow helium gas cryostat was used
for controlling the temperature of the sample.

The SXRD data were measured by the rotation method,
employing a
LAMBDA 7.5M area detector by X-Spectrum \cite{pennicard2012a}.
Frames of width of 0.1 deg and exposure time 0.1 s
were collected during continuous rotation over 364 deg.
At each temperature, the raw data were binned
towards a single run of 363 frames of 1 deg wide.
The software CrysAlisPro \cite{crysalis} was used for
the computation of undistorted views of reciprocal space.
Two complete data sets of SXRD data were
obtained in this way.
The SXRD-160 data for a sample temperature of 160 K,
and the SXRD-30 data for a sample temperature
of 30 K.
In a similar way, the data from Ref. \cite{ramakrishnan2022a}
are designated SXRD-250 and SXRD-70 data.

Data processing of the binned runs was performed
with the EVAL15 software suite \cite{schreursamm2010a}.
Indexing and integration resulted in values for the
lattice parameters and modulation wave vector,
as well as a list of Bragg reflections with their
integrated intensities.
Absorption correction and
scaling were computed for each data set by the
software SADABS \cite{sheldrick2008}.
These corrections are obtained through the
comparison of intensities of equivalent reflections.
The correction thus depends on the point symmetry
that is assumed to be valid for the SXRD data.
Here, the problem arises that different symmetries
lead to comparable agreements between equivalent
reflections, with $R_{int}\approx 3\%$ for main
reflections, and even slightly higher for the
tetragonal phase at 160 K (see Table S1 in the
supplemental material \cite{eual4suppmat2024a}).
This can be explained by twinning of the crystal
in the CDW phase, as it is likely to occur at
the CDW phase transition, if this transition
leads to a lowering of the point group.
For equal volumes of twin domains the diffraction
data of EuAl$_4$ would have $4/mmm$ point symmetry,
irrespective of the symmetry of the crystal
structure in the CDW state.

Refinements of the crystal structures was done using
the software package Jana2020 \cite{petri2024a}.
Details of the experiment, data processing
and crystallographic information are provided
in Table \ref{tab:eual4_cdw_crystalinfo}
and in the supplemental material \cite{eual4suppmat2024a}.
\begin{table}
\caption{\label{tab:eual4_cdw_crystalinfo}%
Crystallographic data of
EuAl$_4$ at 160 K (periodic phase) and 30 K (CDW phase).}
\begin{ruledtabular}
\begin{tabular}{ccc}
Temperature (K)                & 160       & 30 \\
Crystal system                   & Tetragonal & Orthorhombic \\
(Super-)space group    & $I4/mmm$  & $F222(0\,0\,\sigma)00s$ \\
 No. \cite{stokesht2011a}   & 139 & {22.1.17.2}  \\
$a$ (\AA{})                       &4.3922(1)    & 6.2056(1)  \\
$b$ (\AA{})                       &4.3922        & 6.2055(1)  \\
$c$ (\AA{})                       &11.1707(3)  & 11.1630(2)  \\
Volume (\AA{}$^3$)         & 215.50(1)   & 429.88(2)  \\
Wave vector $\mathbf{q}$       & --& 0.1743(1)$\,\mathbf{c}^{*}$  \\
$Z$                                  & 2 & 4 \\
Wavelength (\AA{})           & 0.50000 & 0.50000  \\
Detector distance (mm)      &95 & 95  \\
$\chi$-offset (deg)             &-60 & -60 \\
Rotation per frame (deg)    & 1 & 1  \\
$(\sin(\theta)/\lambda)_{max}$ (\AA{}$^{-1}$) &0.719972 & 0.720842 \\
Absorption coefficient, \\
 $\mu$ (mm$^{-1}$)   & 5.875 & 5.890  \\
T$_{min}$, T$_{max}$ & 0.7057, 0.8618 & 0.6794, 0.8618  \\
Criterion of observability & $I>3\sigma(I)$ & $I>3\sigma(I)$ \\
\multicolumn{2}{l}{No. of reflections measured,} \\
$(m = 0)$  & 1796 & 1144  \\
$(m = 1)$  & -       & 2316 \\
$(m = 2)$  & -       & 2322 \\
Point group for averaging & $4/mmm$ & $mmm$ \\
No. of unique reflections,   \\
$(m = 0)$ (obs/all) & 127/127 & 207/207  \\
$(m = 1)$ (obs/all) & --          & 345/380  \\
$(m = 2)$ (obs/all) & --          & 31/394    \\
$R_{int}$ (obs/all)         &0.0386/0.0386 & 0.0407/0.0414 \\
$R_{int}(m = 0)$ (obs/all)  &0.0386/0.0386 & 0.0303/0.0303 \\
$R_{int}(m = 1)$ (obs/all)  &--            & 0.1056/0.1063 \\
$R_{int}(m = 2)$ (obs/all)  &--            & 0.1653/0.2325 \\
No. of parameters &9 &40  \\
$R_{F }(m = 0)$  (obs) &0.0193 &0.0167  \\
$R_{F }(m = 1)$  (obs) &-          &0.0416  \\
$R_{F }(m = 2)$  (obs) &-          &0.0566  \\
$wR_{F }(m = 0)$ (all) &0.0230 &0.0204  \\
$wR_{F }(m = 1)$ (all) &-         &0.0477  \\
$wR_{F }(m = 2)$ (all) &-         &0.1921  \\
$wR_{F }$ (all) (all refl.)   &0.0230 &0.0317  \\
GoF (obs/all)                &1.71/1.71 &1.37/1.11 \\
$\Delta\rho_{min}$, $\Delta\rho_{max}$(e \AA{}$^{-3}$) &
 -1.35, 1.39 & -4.6, 5.24 \\
\end{tabular}
\end{ruledtabular}
\end{table}
%
%


\section{\label{sec:eual4_symmetry_structure}%
Symmetry and crystal structure}

The diffraction data at 160 K (SXRD-160 data) confirm
the BaAl$_4$ structure type with space group $I4/mmm$
for EuAl$_4$ \cite{shangt2025a}.
Structure refinements led to an excellent
fit to the SXRD-160 data with $R_{F}$ = 0.0193
(Table \ref{tab:eual4_cdw_crystalinfo}).
Parameters of the three crystallographically
independent atoms---Eu, Al1 and Al2---are
given in Table S6 in the supplemental material
\cite{eual4suppmat2024a}.

\begin{figure}
\includegraphics[width=78mm]{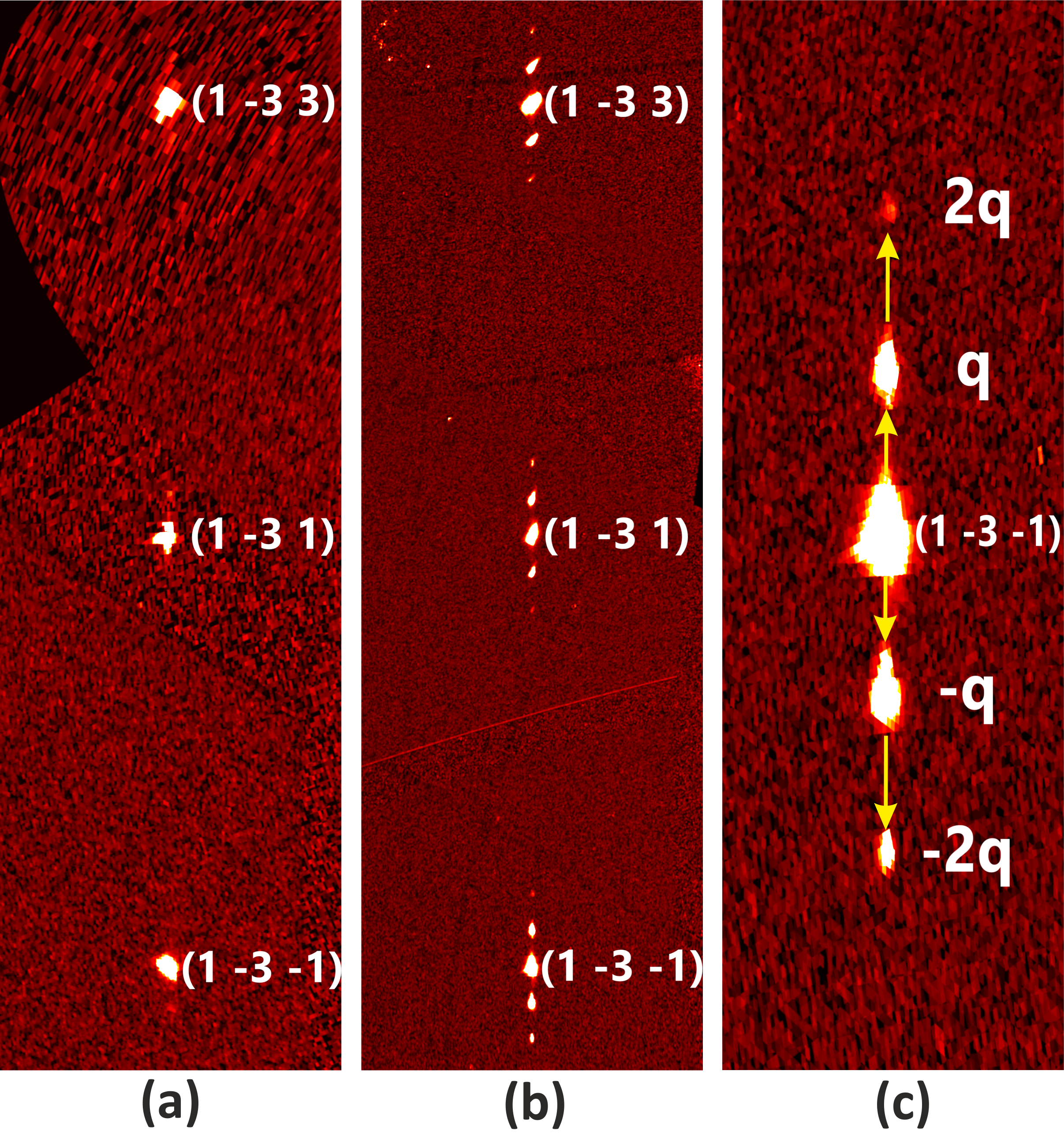}
\caption{\label{fig:eual4_unwarp}%
Undistorted view of the $(1\,k\,l)$  reciprocal lattice
plane of
(a) the SXRD-160 data, and
(b) the SXRD-30 data.
(c) Enlarged view of panel (b) about reflection
$(1\, {-}3\, -1)$, with clearly visible
first-order and second-order satellite reflections.
Images have been generated from the measured data
by the software CrysAlisPro \cite{crysalis}.
Dark bands are due to gaps (insensitive area) between
the active modules of the Lambda 7.5M area detector.
}
\end{figure}

Cooling of the crystal towards 30 K resulted
in the appearance of satellite reflections in
the SXRD \cite{shimomura2019a}.
The observation of
up to third-order satellite reflections has
been reported in the literature,
but only first-order satellite reflections
have been used for structural analysis
\cite{shimomura2019a,ramakrishnan2022a,korshunov2024a}.
Presently, the SXRD-30 data comprise both first-order
and second-order satellite reflections
(Figure \ref{fig:eual4_unwarp} and Table S1 in
\cite{eual4suppmat2024a}).

\begin{figure}
\includegraphics[width=78mm]{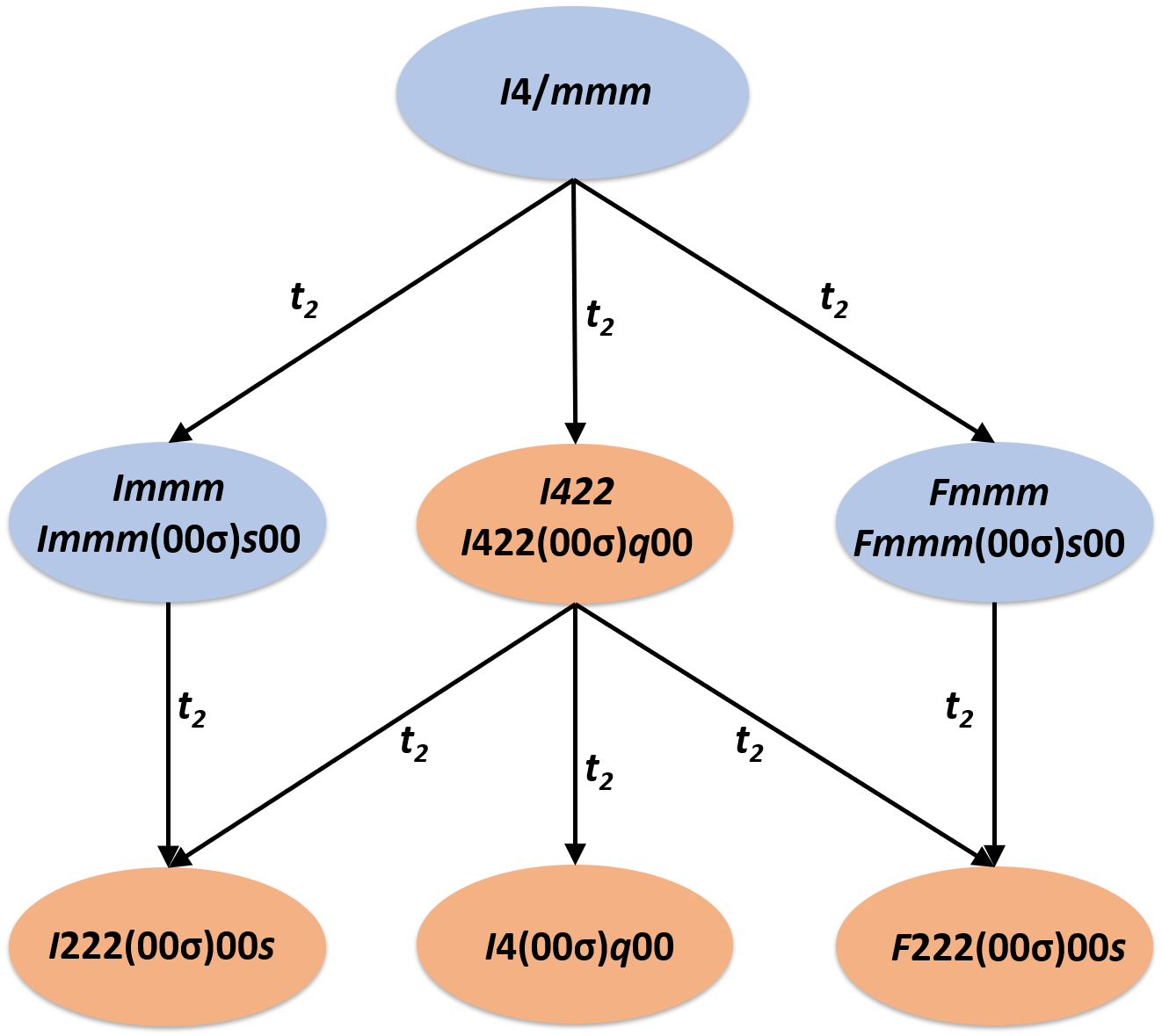}
\caption{\label{fig:eual4_ssg_flowchart}%
Flow chart showing the relations between
tetragonal and orthorhombic space groups and
superspace groups, where the basic-structure
space group is a subgroup of $I4/mmm$.
Furthermore, the acentric orthorhombic
superspace groups (bottom row) are subgroups
of the superspace groups in the middle row.
''t2'' indicates that the point group is a
subgroup of index two.
Centrosymmetric superspace groups are in light blue
and acentric superspace groups are in light brown color.}
\end{figure}

The incommensurately modulated crystal structure of
EuAl$_4$ in its CDW state
is described within the superspace approach
\cite{vansmaalen2012a,janssent2018a}.
In principle, any superspace group is a candidate
symmetry for the CDW phase, that is based on a
modulation wave vector along $\mathbf{c}^{*}$
and a basic-structure space group that is a
''translationsgleiche'' subgroup of $I4/mmm$.
A total of $63$ superspace groups exist with these
properties \cite{ramakrishnan2024a}.
Previous analyses of SXRD data of up to first-order
satellites have shown that six of these $63$
superspace groups allow for a structure model
that may describe the SXRD data well
\cite{ramakrishnan2022a,korshunov2024a,ramakrishnan2024a}.
Present refinements confirm this finding for the
SXRD-30 data (Tables S2 and S3 in \cite{eual4suppmat2024a}).
These symmetries comprise two acentric tetragonal superspace
groups, two acentric orthorhombic superspace groups
and two centrosymmetric orthorhombic superspace
groups (Fig. \ref{fig:eual4_ssg_flowchart} and
Table \ref{tab:eu30_ref_ssgave}).
\begin{table*}
\centering
\caption{\label{tab:eu30_ref_ssgave}%
Structure refinements against the SXRD-30 data of EuAl$_4$
in its incommensurate CDW phase at $T=30$ K.
Three types of structure model are considered
for each of six superspace symmetries.
''Basic structure'' refers to a refinement of the
basic structure against main reflections $(m=0)$ only;
''Displacement modulation'' adds to the basic structure
first-order harmonic modulation parameters of
displacement modulation;
''Displacement + ADP modulation'' adds to ''displacement
modulation'' second-order harmonic modulation parameters
for the anisotropic displacement parameters (ADPs).
SXRD-30 data have been processed according to
the centrosymmetric point group of the superspace
group (Table S1 in \protect\cite{eual4suppmat2024a}).
Criterion of observability is $I>3\sigma(I)$.
$n_{par}$ is the number of refined parameters.
'{*}' indicates refinements with a $\Delta F$ problem
(see text).
}
\begin{ruledtabular}
\begin{tabular}{lcccccc}
&   &   &$R_F^{obs}$ & $R_F^{obs}$ &$R_F^{obs}$ & $R_F^{obs}$     \\
SSG  & Averaging  & $n_{par}$ & (Overall) &($m=0$) & ($m=1$) & ($m=2$)   \\
&   &   & (\%) & (\%) & (\%) &(\%)  \\
\hline
\multicolumn{2}{c}{\textbf{Basic structure}}   \\
\hline
$I422(0\,0\,\sigma)q00$  & $4/mmm$ & 9  & 1.79  & 1.79 & -  & -  \\
$I4(0\,0\,\sigma)q00$    & $4/m$   & 11 & 1.68  & 1.68 & -  & -  \\
$Immm(0\,0\,\sigma)s00$  & $Immm$  & 9  & 3.24  & 3.24 & -  & -  \\
$I222(0\,0\,\sigma)00s$  & $Immm$  & 9  & 3.24  & 3.24 & -  & -  \\
$Fmmm(0\,0\,\sigma)s00$  & $Fmmm$  & 9  & 1.81  & 1.81 & -  & -  \\
$F222(0\,0\,\sigma)00s$  & $Fmmm$  & 9  & 1.81  & 1.81 & -  & -  \\
\hline
\multicolumn{2}{c}{\textbf{Displacement Modulation }}   \\
\hline
$I422(0\,0\,\sigma)q00$  & $4/mmm$ & 14* & 3.37 & 2.12 & 6.12 & 4.59  \\
$I4(0\,0\,\sigma)q00$    & $4/m$   & 21* & 3.31 & 1.94 & 6.03 & 4.74  \\
$Immm(0\,0\,\sigma)s00$  & $Immm$  & 14* & 4.81 & 3.36 & 7.37 & 20.31 \\
$I222(0\,0\,\sigma)00s$  & $Immm$  & 19* & 4.28 & 3.31 & 6.39 & 3.85  \\
$Fmmm(0\,0\,\sigma)s00$  & $Fmmm$  & 15* & 4.14 & 2.21 & 7.58 & 23.63 \\
$F222(0\,0\,\sigma)00s$  & $Fmmm$  & 19* & 3.5  & 2.05 & 6.5  & 4.9   \\
\hline
\multicolumn{2}{c}{\textbf{Displacement + ADP Modulation}}   \\
\hline
$I422(0\,0\,\sigma)q00$  & $4/mmm$ & 24  & 3.37 & 2.1  & 6.01 & 7.55  \\
$I4(0\,0\,\sigma)q00$    & $4/m$   & 41  & 2.93 & 1.79 & 5.05 & 8.23  \\
$Immm(0\,0\,\sigma)s00$  & $Immm$  & 29  & 4    & 3.32 & 5.13 & 13.09 \\
$I222(0\,0\,\sigma)00s$  & $Immm$  & 40  & 3.63 & 3.17 & 4.57 & 5.19  \\
$Fmmm(0\,0\,\sigma)s00$  & $Fmmm$  & 28  & 2.86 & 1.81 & 4.78 & 12.05 \\
$F222(0\,0\,\sigma)00s$  & $Fmmm$  & 40  & 2.51 & 1.67 & 4.16 & 5.66  \\
\end{tabular}
\end{ruledtabular}
\end{table*}

Goal of the present analysis is to determine the
symmetry and crystal structure of the CDW phase.
First, it is noticed that the lattice remains of
tetragonal metric in the CDW phase.
No splitting nor broadening has been observed
of Bragg reflections (Fig. \ref{fig:eual4_unwarp}),
in agreement with the literature
\cite{shimomura2019a,ramakrishnan2022a,korshunov2024a,yangfz2024a}.
This observation indicates that the lattice symmetry
remains tetragonal within experimental error.
However, as shown below, the symmetry of the
crystal structure is lowered to acentric orthorhombic
in the CDW phase.

The first model to be considered is the basic structure
(no modulation).
Except for $I4$, the basic-structure coordinates
remain equal to those of $I4/mmm$, where there
exists only one refinable coordinate, $z$[Al2]
(Table S6 in \cite{eual4suppmat2024a}).
Special case is the symmetry $F222(0\,0\,\sigma)00s$,
in which the Al1 site splits into two sites,
denoted as Al1a and Al1b.
Since Al1, Al1a and Al1b do not incorporate
refinable coordinates, the symmetry of the
basic structure remains $I4/mmm$ here too.
The orthorhombic symmetries do allow for more
independent ADP parameters.
However, lower symmetry for the ADPs is unlikely
for an undistorted basic structure.
Therefore, the basic-structure ADPs were restricted
to follow $I4/mmm$ symmetry in all refinements.
An excellent fit of the main reflections of the
SXRD-30 data has been obtained
(Table \ref{tab:eu30_ref_ssgave}).
These results imply that the basic structure
of the incommensurate CDW phase remains tetragonal,
and that any distortions will be due to the
modulation wave.
The less good agreement for the $I$-centered
orthorhombic symmetry is attributed to the less
than optimal performance of the SADABS refinement
for this symmetry.

The second model describes the displacive modulation
by first-order harmonic parameters.
Although a reasonable fit to the SXRD-30 data is obtained
for all six symmetries,
except for the fit to the second-order satellite
reflections for models with centrosymmetric symmetries
(Table \ref{tab:eu30_ref_ssgave}),
several problems remain.
Firstly, the fit to the main reflections is less
good than in the basic-structure refinement.
However, this fit should improve upon the introduction
of the correct
model for the incommensurate modulation.
Apparently, something is lacking for this model.
In this respect, it is noticed that the standard procedure
would be the introduction of second-order harmonic
parameters for the displacement modulation,
if the SXRD data
contain second-order satellite reflections,
as the SXRD-30 data do.
However, the second problem with the simple
modulated structure is that second-order
satellite reflections are calculated too strong,
\textit{i.e.}
\begin{equation}
\Delta F(h\,k\,l\,m) =
\left(F_{obs}(h\,k\,l\,m) - F_{cal}(h\,k\,l\,m)\right) < 0
\label{eqn:eual4_delta_f}
\end{equation}
for most second-order satellites ($m=\pm 2$).
We have called this the $\Delta F$ problem,
and it is found for all symmetries
(Table \ref{tab:eu30_ref_ssgave}).
Non-zero values for the second-order harmonic parameters
for the displacive modulation lead to even
higher calculated structure factor amplitudes
$F_{cal}(h\,k\,l\,m)$ for the second-order satellites,
aggravating the $\Delta F$ problem.
Indeed, refinements of the displacement modulation
up to second-order harmonic parameters leads to
insignificant or zero values for the latter.

Instead of modifying the displacement modulation
functions, modulations can be introduced for
the ADPs.
First-order harmonic parameters for the modulation
of the ADPs did not lead to an improved fit to
the SXRD-30 data, while those parameters refined
to values smaller than their standard uncertainties.
Therefore, we have restricted to zero the values
of these parameters in the remaining analysis.
The third model that we have considered is
the combination of first-order harmonic
parameters for displacive modulation with
second-order harmonic parameters for the
modulation of ADPs.
A clearly improved fit to the SXRD-30 data
was thus obtained for Five out of six superspace
symmetries (Table \ref{tab:eu30_ref_ssgave}),
while the $\Delta F$-problem is resolved.
However, we notice that the best fit with
$R_{F}^{obs}(\mathrm{overall})$ = 2.51\% is obtained
for the non-centrosymmetric orthorhombic
superspace group $F222(0\,0\,\sigma)00s$.
Furthermore, the two tetragonal superspace
groups lead to higher values for $R_{F}^{obs}(m=0)$
than in the basic-structure refinement;
an additional argument against these symmetries.
Both centrosymmetric orthorhombic superspace
groups---previously proposed on the basis of
SXRD data without second-order satellites
\cite{ramakrishnan2022a,korshunov2024a,ramakrishnan2024a}---%
can be excluded, because of the poor fit to
the second-order satellite reflections.

%
%
An alternate approach is to use the same
reflection list for all refinements.
This list then should be based on the common
point symmetry of all six superspace groups, which is
$\mathbf{c}$-unique monoclinic symmetry.
Accordingly, SXRD-30 data have been processed within
$2/m$ point symmetry with, apparently, similar results
as for processing within the other symmetries
(Table S1 in \cite{eual4suppmat2024a}).
However, refinements show a less good fit
to the main reflections, indicating that the
computation of the absorption correction and
other scalings was less successful for $2/m$
symmetry than for the higher symmetries,
probably because of the lower redundancy for
$2/m$ symmetry.
Nevertheless, with these data,
$F222(0\,0\,\sigma)00s$ is marginally preferred
over $I222(0\,0\,\sigma)00s$
(Table S4 in \cite{eual4suppmat2024a}).
We attribute this very small difference to the almost
perfect twinning (all domains of equal volume)
of the crystal while measuring the SXRD-30 data.
With inversion twins restricted to be equal,
refined twin volumes are
0.253(2):0.247 for $I222(0\,0\,\sigma)00s$ and
0.257(2):0.243 for $F222(0\,0\,\sigma)00s$ symmetry.
The SXRD-70 data were measured on a different
crystal that appeared to be twinned with
twin volumes $0.29 :\: 0.21$.
This deviation from perfect twinning has appeared
to be sufficient for a more clear preference of
$F222(0\,0\,\sigma)00s$ over $I222(0\,0\,\sigma)00s$
symmetry, despite the lack of second-order
satellites
(Table \ref{tab:eu70_ref_monoave}
and Table S5 in \cite{eual4suppmat2024a}).
\begin{table*}
\caption{\label{tab:eu70_ref_monoave}%
Structure refinements against the SXRD-70 data of
EuAl$_4$ in its incommensurate CDW phase at $T = 70$ K.
Data from \protect\cite{ramakrishnan2022a}.
The same types of structure model and the same
symmetries are considered as in
Table \protect\ref{tab:eu30_ref_ssgave}.
The SXRD-70 data have been processed according to
the centrosymmetric point group $2/m$
($\mathbf{c}$ unique; Table S1 in \cite{eual4suppmat2024a}).
Criterion of observability is $I>3\sigma(I)$.
$n_{par}$ is the number of refined parameters.
A $\Delta F$-problem does not exist, because
second-order satellites are not part of the SXRD-70 data.}
\begin{ruledtabular}
\begin{tabular}{lcccccc}
 &   &   &$R_F^{obs}$ & $R_F^{obs}$ &$R_F^{obs}$ & $R_F^{all}$     \\
SSG  & Averaging  & $n_{par}$ & (Overall) &($m=0$) & ($m=1$) & (all)   \\
 &   &   & (\%) & (\%) & (\%) &(\%)  \\
\hline
\multicolumn{2}{c}{\textbf{Basic structure}}      \\
\hline
$I422(0\,0\,\sigma)q00$     & $2/m$  & 9     & 2.05  & 2.05  & -       & 2.05   \\
$I4(0\,0\,\sigma)q00$       & $2/m$  & 11    & 2.03  & 2.03  & -       & 2.03   \\
$Immm(0\,0\,\sigma)s00$ & $2/m$   & 9      & 2.05  & 2.05  & -       & 2.05   \\
$I222(0\,0\,\sigma)00s$      & $2/m$   & 9      & 2.05  & 2.05  & -       & 2.05 \\
$Fmmm(0\,0\,\sigma)s00$ & $2/m$  & 9      & 2.05  & 2.05  & -       & 2.05   \\
$F222(0\,0\,\sigma)00s$    & $2/m$   & 9      & 2.05  & 2.05  & -       & 2.05   \\
\hline
\multicolumn{2}{c}{\textbf{Displacement Modulation}}     \\
\hline
$I422(0\,0\,\sigma)q00$      & $2/m$      & 14    & 3.21  & 2.06  & 6.22  & 3.38  \\
$I4(0\,0\,\sigma)q00$          & $2/m$      & 21    & 3.19  & 2.04  & 6.17  & 3.38  \\
$Immm(0\,0\,\sigma)s00$     & $2/m$     & 16    & 3.2   & 2.13  & 5.99  & 3.36 \\
$I222(0\,0\,\sigma)00s$      & $2/m$      & 21    & 3.15  & 2.09  & 5.92  & 3.33   \\
$Fmmm(0\,0\,\sigma)s00$ & $2/m$       & 15    & 3.14  & 2.16  & 5.7   & 3.28  \\
$F222(0\,0\,\sigma)00s$      & $2/m$     & 19     & 3.2   & 2.06  & 6.18  & 3.37   \\
\hline
\multicolumn{2}{c}{\textbf{Displacement + ADP Modulation}} \\
\hline
$I422(0\,0\,\sigma)q00$     & $2/m$      & 24     & 3.01  & 2.07  & 5.45  & 3.22 \\
$I4(0\,0\,\sigma)q00$        & $2/m$      & 41    & 2.79  & 1.85  & 5.22  & 2.98 \\
$Immm(0\,0\,\sigma)s00$  & $2/m$      & 31  & 2.74  & 1.97   & 4.73  & 3.07  \\
$I222(0\,0\,\sigma)00s$     & $2/m$     & 41    & 2.46  & 1.65  & 4.56  & 2.64   \\
$Fmmm(0\,0\,\sigma)s00$ & $2/m$      & 28  & 2.76  & 2.02    & 4.7   & 2.96  \\
$F222(0\,0\,\sigma)00s$    & $2/m$     & 40    & 2.41  & 1.67  & 4.34  & 2.59   \\
\end{tabular}
\end{ruledtabular}
\end{table*}

\section{\label{sec:eual4_discussion}%
Discussion}

The symmetry of the CDW state of EuAl$_4$ has been
found as non-centrosymmetric orthorhombic with
superspace group $F222(0\,0\,\sigma)00s$.
This superspace group is a subgroup of
centrosymmetric $Fmmm(0\,0\,\sigma)s00$, which
was previously proposed as symmetry
\cite{ramakrishnan2022a}.
Alternatively, the other centrosymmetric
orthorhombic superspace group,
$Immm(0\,0\,\sigma)s00$, was proposed as symmetry
\cite{korshunov2024a}
(Fig. \ref{fig:eual4_basic_unit_cell}).
Here, the $I$-centered orthorhombic groups
preserve the twofold axes along the coordinate
axes of $I4/mmm$, and the $F$-centered orthorhombic
groups preserve the diagonal twofold axes of
$I4/mmm$.
We believe that the distinction could be made
between all six possible superspace groups in
Table \ref{tab:eu30_ref_ssgave}, because the
present SXRD-30 diffraction data is an extensive
data set containing second-order satellite
reflections.
Although observed, previous structural analysis
was based on SXRD data containing only first-order
satellite reflections
\cite{ramakrishnan2022a,korshunov2024a}.
The finding of acentric symmetry is in agreement
with the isostructural, non-magnetic compound SrAl$_4$,
which features a stronger CDW with stronger
second-order satellites, that are also described by
the acentric superspace group
$F222(0\,0\,\sigma)00s$ \cite{ramakrishnan2024a}.
The absence of inversion symmetry immediately implies
that the previously reported skyrmion lattices
could be stabilized by the DM interaction and do not
need more exotic mechanisms
\cite{muhlbaue2009a,takagir2022a,gen2023a,nih2024a}.

Whereas the choice of $F222(0\,0\,\sigma)00s$
is principally based on the quality of the fit
to the SXRD data of the CDW phase, the structure
refinement also leads to a structure model for
the CDW phase of EuAl$_4$.
First, it is to be noted that the basic structure
remains $I4/mmm$ (Table S6 in \cite{eual4suppmat2024a}).
Only ADP parameters could deviate from this symmetry,
but they were fixed to the tetragonal symmetry,
because a lower symmetry of the ADP parameters is
unlikely for a crystal structure that lacks distortions.
The lower symmetry, including the loss of inversion
symmetry, thus is entirely due to the
symmetry of the CDW modulation.
Such a phenomenon has been observed before, for example
for Mo$_2$S$_3$ \cite{schuttewj1993b},
Sm$_2$Ru$_3$Ge$_5$ \cite{bugaris2017a} and
Gd$_2$Os$_3$Si$_5$ \cite{vikash2024a}.
Presently, the CDW modulation represents a transverse
wave (Table S7 in \cite{eual4suppmat2024a}),
in agreement with \cite{yangfz2024a}.

\begin{figure}
\includegraphics[width=70mm]{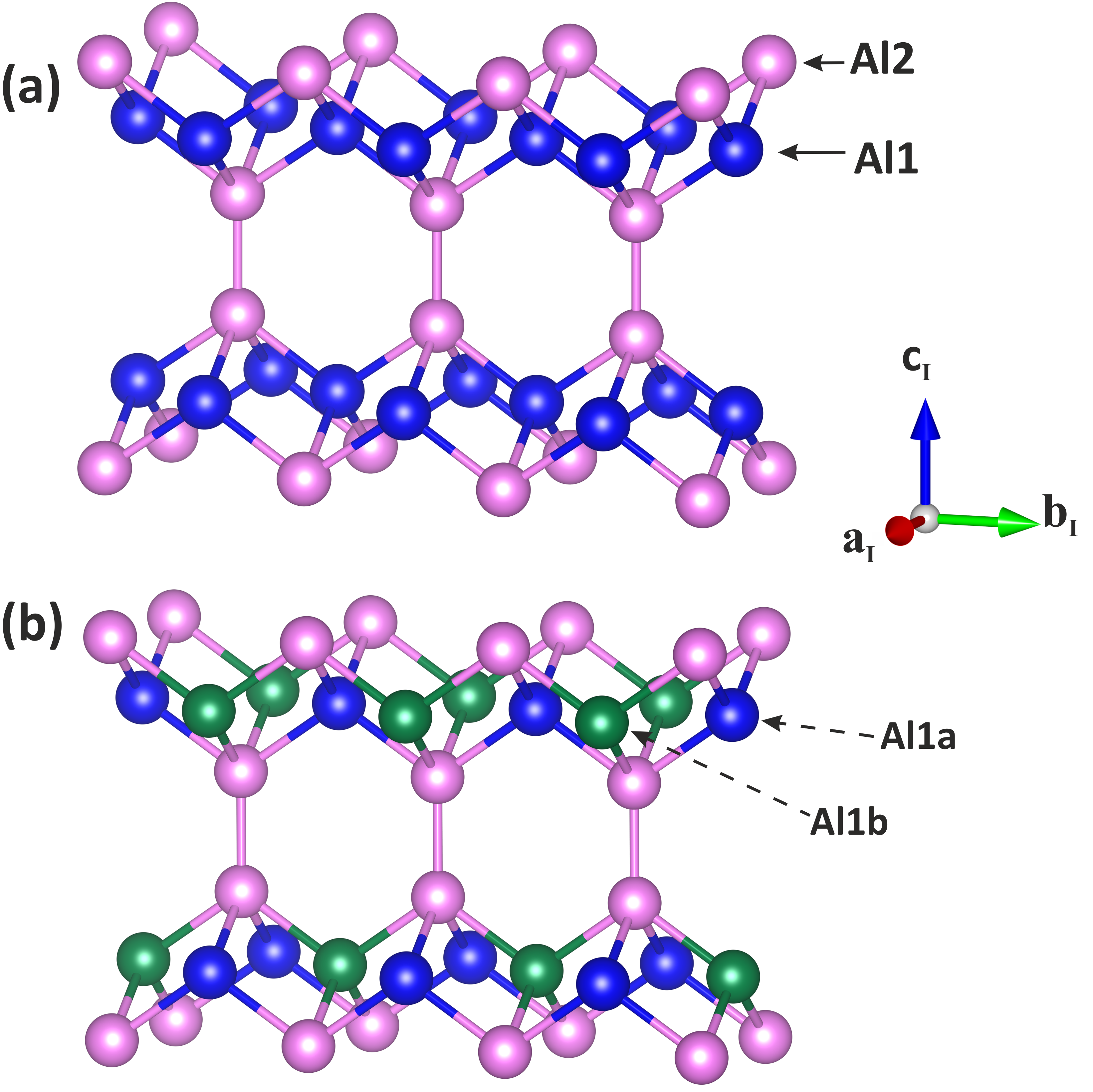}
\caption{\label{fig:eual4_layer}%
Perspective view of layers of aluminum atoms in EuAl$_4$.
(a) Layer of Al atoms in the periodic phase, where the
shortest distances are within the Al2-Al2 dumbbells
with $d$[Al2--Al2] = 2.562(3) \AA{} at 160 K.
Next shortest distances are $d$[Al2--Al1] = 2.666(2) \AA{}
and $d$[Al1--Al1] = 3.1058(1) \AA{}.
(b) The CDW phase featuring Al1a and Al1b independent
atoms in the basic structure of symmetry $F222$.
Basic-structure distances in the modulated CDW phase at 30K are
$d$[Al2--Al2] = 2.566(2) \AA{},
$d$[Al2--Al1a] = $d$[Al2--Al1b] = 2.662(1) \AA{}
and $d$[Al1a--Al1b] = 3.1027 \AA{}.
The modulation of these distances is given as t-plots
in Fig. \protect\ref{fig:eual4_t_plot_al}.}
\end{figure}

\begin{figure*}
\includegraphics[width=170mm]{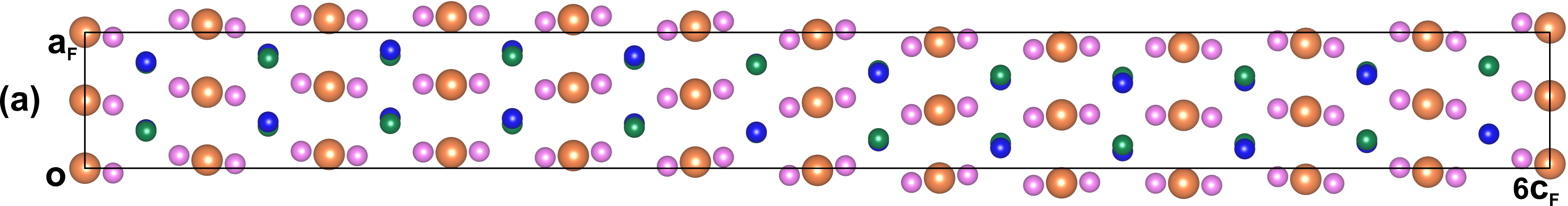}
\par\vspace{4mm}
\includegraphics[width=170mm]{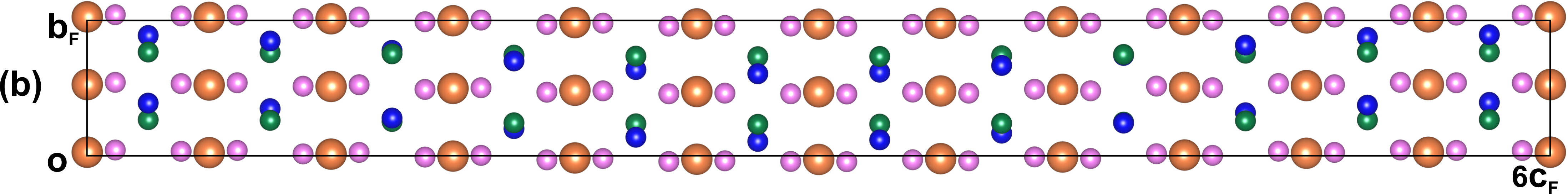}
\caption{\label{fig:eual4_6fold_modulation}%
Incommensurately modulated crystal structure of EuAl$_4$
for symmetry $F222(0\,0\,\sigma)00s$
(Table \protect\ref{tab:eual4_cdw_crystalinfo}).
Black lines represent the boundary of $1\times 1\times 6$
unit cells of the basic structure.
All atoms are at their modulated (displaced) positions.
Eu in orange, Al1a in blue, Al1b in green and Al2 in pink.
(a) Projection along $\mathbf{b}_{F}$, showing
displacements along $\mathbf{a}_{F}$.
(b) Projection along $\mathbf{a}_{F}$, showing
displacements along $\mathbf{b}_{F}$.
Atomic displacements have been magnified by
a factor of $5$.
Displacements are zero along $\mathbf{c}_{F}$.
}
\end{figure*}

\begin{figure}
\includegraphics[width=70mm]{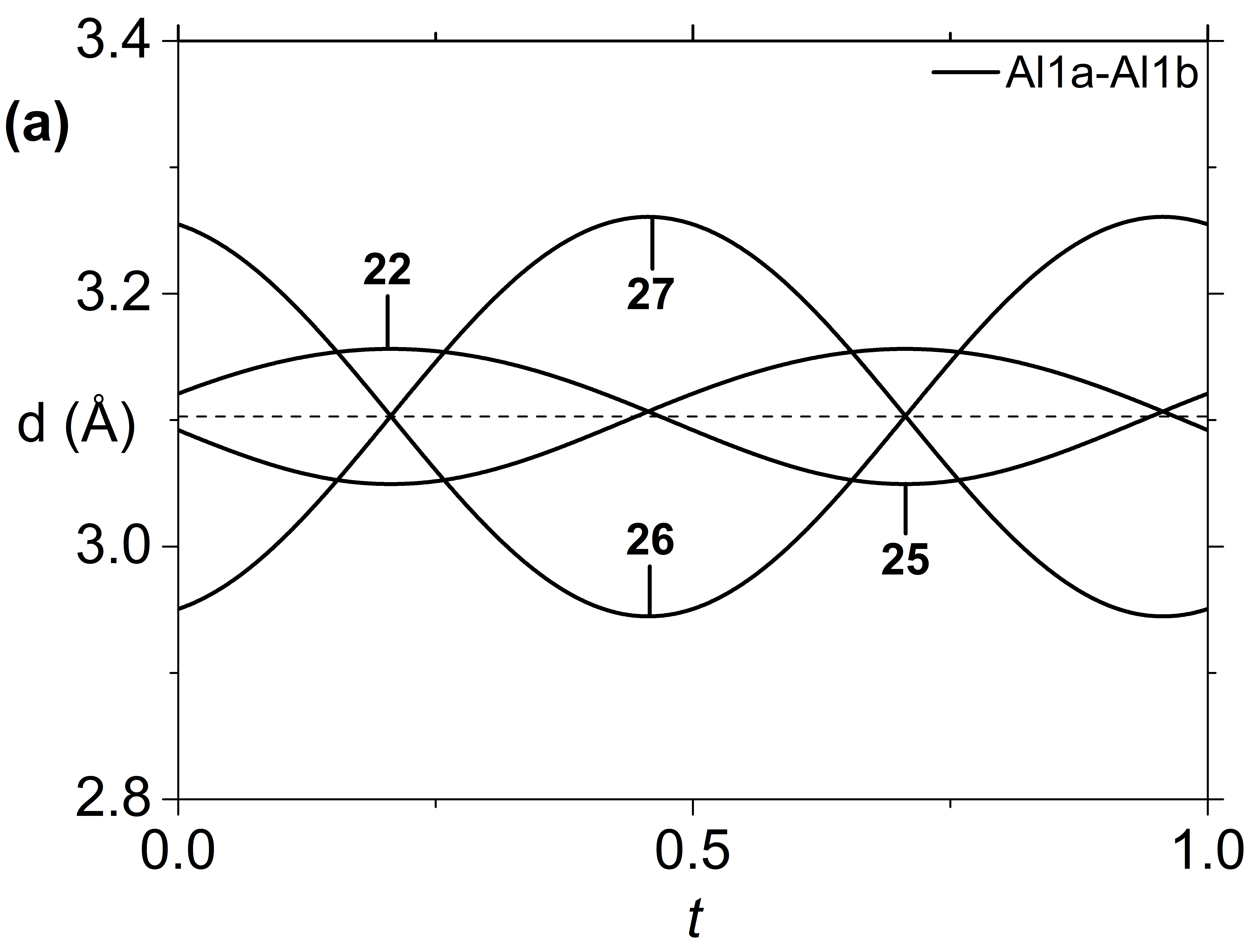}
\includegraphics[width=70mm]{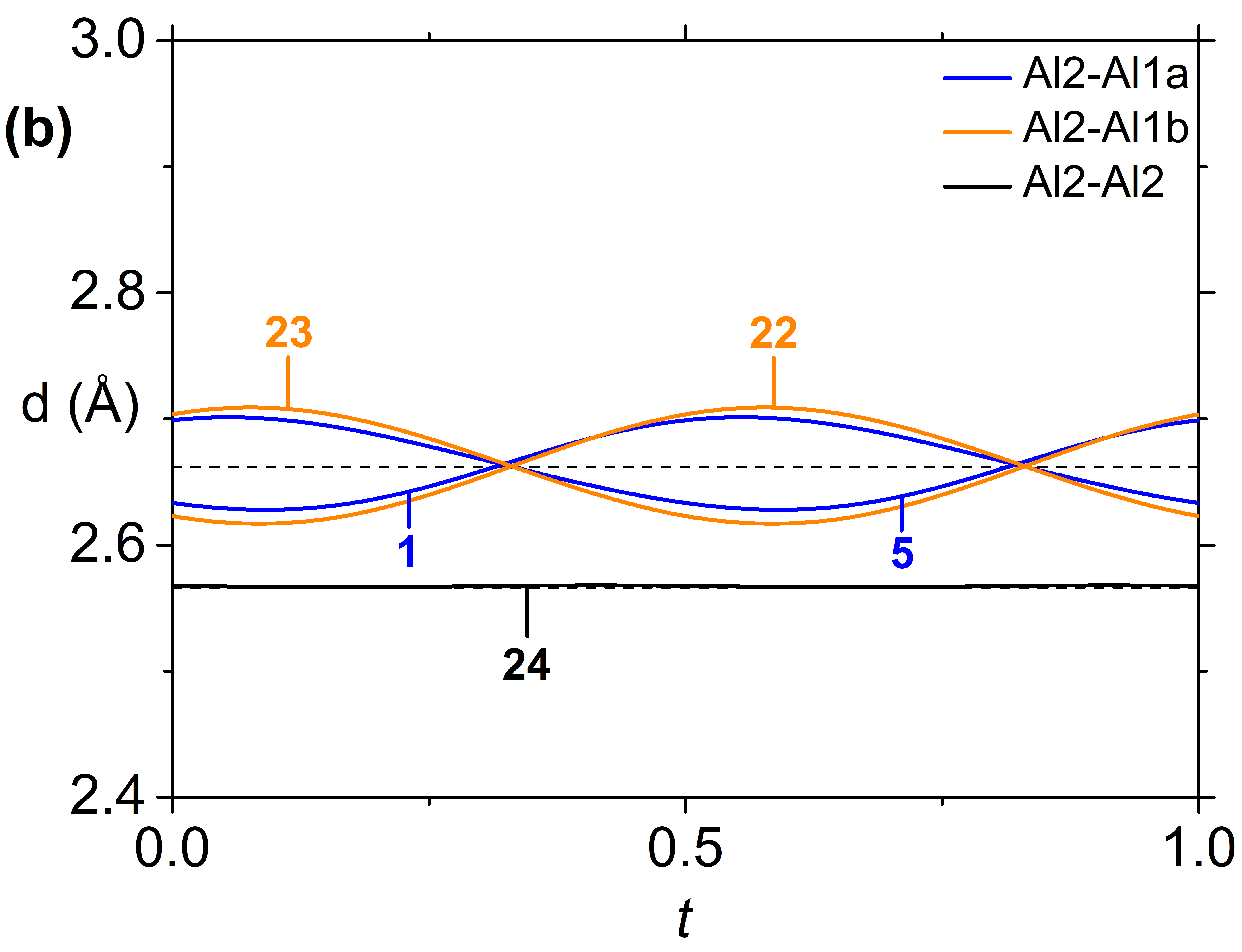}
\caption{\label{fig:eual4_t_plot_al}%
$t$-Plots of interatomic distances (\AA{}) in the
CDW phase of EuAl$_4$ at 30 K.
(a) $d$[Al1a--Al1b] with Al1a as central atom, and
(b) $d$[Al2--Al1a], $d$[Al2--Al1b] and $d$[Al2--Al2]
with Al2 as central atom.
Basic structure distances are indicated by the dashed
horizontal lines.
$t$-Plots display variation in atomic parameters
such as distance, position as a function of the phase $t$ of the
modulation wave \cite{vansmaalen2012a}.
Here, each value of $t$
gives the distances from a central atom
towards its neighboring atoms.
The number on each curve is the number of the symmetry
operator that is applied to the second atom of the
bond pair.
Symmetry operators are listed in Table S9
in the supplementary material
\protect\cite{eual4suppmat2024a}.}
\end{figure}

A second point to notice is that the two centrosymmetric
orthorhombic superspace groups as well as the
acentric tetragonal superspace groups are not subgroups
of any superspace group based on $I4/mmm$.
On the other hand, $F222(0\,0\,\sigma)00s$ is a subgroup
of both $F422(0\,0\,\sigma)q00$  and, previously proposed,
$Fmmm(0\,0\,\sigma)s00$
(Fig. \ref{fig:eual4_ssg_flowchart}).
Indeed, the present acentric structure model can be
considered as a distortion of the
$Fmmm(0\,0\,\sigma)s00$ model
(Compare the discussion in Section S6
in the Supplemental Material \cite{eual4suppmat2024a}).
Initially, it appears that all atoms exhibit similar
displacements along $\mathbf{a}_{F}$
(Figs. \ref{fig:eual4_layer} and
\ref{fig:eual4_6fold_modulation}(a), and
Table S7 and Figs. S1–S3 in \cite{eual4suppmat2024a}).
However, major effect is the unequal displacement
modulations of the Al1a and Al1b atoms into the
direction of $\mathbf{b}_{F}$
(Figs. \ref{fig:eual4_layer} and
\ref{fig:eual4_6fold_modulation}(b), and
Table S7 and Figs. S1–S3 in \cite{eual4suppmat2024a}).
$t$-Plots reveal that the major modulation of
interatomic distances is between Al1 atoms and
of secondary importance between Al2 and Al1 atoms
(Fig. \ref{fig:eual4_t_plot_al}).
This finding is in agreement with the
$Fmmm(0\,0\,\sigma)s00$ model \cite{ramakrishnan2022a},
and it confirms that the CDW resides on the aluminum
layers \cite{kobata2016a,kaneko2021a}.
The acentric distortion of the CDW
according to $F222(0\,0\,\sigma)00s$
appears to be
in agreement with the structure proposed by
Ni \textit{et al.} \cite{nih2024a} on the basis
of 4D-STEM data, although in the latter report
no symmetry group was given
(see the discussion in Section S7 of the
Supplemental Material \cite{eual4suppmat2024a}).

Korshunov \textit{et al.} \cite{korshunov2024a}
have proposed for the CDW phase of EuAl$_4$
the centrosymmetric superspace group $Immm(0\,0\,\sigma)s00$.
We have recently found this symmetry for the CDW
phase of EuAl$_2$Ga$_2$ \cite{agarwalh2025a}.
In view of the present results for EuAl$_4$, it
is likely that the true symmetry of EuAl$_2$Ga$_2$
could be $I222(0\,0\,\sigma)00s$, since the analyses
in \cite{korshunov2024a} and \cite{agarwalh2025a}
were based on SXRD data without
second-order satellite reflections,
for which it is difficult to distinguish
acentric from centrosymmetric symmetries.
Also, consideration of the four orthorhombic symmetries
listed in Fig. \ref{fig:eual4_ssg_flowchart} and
Table \ref{tab:eu30_ref_ssgave} shows that they share
the tetragonal basic structure as well as
a modulation that is strongest on the
layers of Al1 atoms
(compare to Section S5 in \cite{eual4suppmat2024a}) \cite{ramakrishnan2022a,ramakrishnan2024a}.
Therefore, it is conceivable that the CDWs can
be stabilized by modulations according to either
of these symmetries.
It is then possible that $F222(0\,0\,\sigma)00s$
is most stable for EuAl$_4$, while
$I222(0\,0\,\sigma)00s$ is most stable for EuAl$_2$Ga$_2$.
It might even be the case that a different symmetry
is achieved for the CDW state of a single compound,
depending on its chemical purity and the
concentration of lattice defects.

A modulation of the ADPs was required, in order
to resolve the $\Delta F$ problem of the second-order
satellite reflections.
Structure refinements have shown that first-order
harmonic modulation parameters were without effect
and refined to value zero, while the second-order
harmonic modulation parameters lead to the good
fit to the SXRD-30 data.
P\'{e}rez-Mato \textit{et al.} \cite{perezmato1991a}
have shown that second-order harmonic modulation
of ADPs reflects the presence of phasons in the
incommensurate modulation wave.
Therefore, we propose that the CDW modulation in EuAl$_4$
and related compounds includes phason disorder.

\section{\label{sec:eual4_conclusions}%
Conclusions}

Broken inversion symmetry of the CDW phase is
established for EuAl$_4$.
Essential experimental information was the
presence of second-order satellites in the
SXRD data set.
The CDW modulation is transverse, like it
was found for isostructural SrAl$_4$ \cite{ramakrishnan2024a},
and it is best described by the acentric orthorhombic
superspace group $F222(0\,0\,\sigma)00s$.
Despite conflicting conclusions,
$F222(0\,0\,\sigma)00s$
appears to be in agreement with all
experimental results on EuAl$_4$
\cite{ramakrishnan2022a,korshunov2024a,sukhanovas2025a,yangfz2024a,nih2024a}.

The CDW is found to reside on the Al--Al network,
in agreement with previous models.
The acentric nature of the CDW modulation
is most clearly established by the unequal
modulations of the Al1a and Al1b atoms.
It should be considered, when considering models
for the stabilization of the skyrmion states
of EuAl$_4$.

\begin{acknowledgments}
Single crystals of EuAl$_4$ were grown by
Kerstin K\"{u}spert at the Laboratory of
Crystallography in Bayreuth.
We thank Kerstin K\"{u}spert and Carsten Paulmann
for their assistance in data collection
at beamline P24.
We acknowledge DESY (Hamburg, Germany),
a member of the Helmholtz Association HGF, for
the provision of experimental facilities.
Beamtime was allocated for proposal I-20231047.
S. R. thanks the Agence Nationale de la Recherche
for support within the project
SUPERNICKEL (Grant No. ANR-21-CE30-0041-04).
This research has been funded by the
Deutsche Forschungsgemeinschaft
(DFG, German Research Foundation) – 406658237.
\end{acknowledgments}

%

\clearpage

\end{document}